\documentclass[sigconf,authorversion]{acmart}
\usepackage{arydshln}
\usepackage{amsmath}
\usepackage{multirow}
\usepackage{listings}
\usepackage{listingsLanguages}
\usepackage{caption,subcaption}
\usepackage{mathtools}
\usepackage[para]{footmisc}

\definecolor{maroon}{cmyk}{0, 0.87, 0.68, 0.32}
\definecolor{halfgray}{gray}{0.55}
\definecolor{ipython_frame}{RGB}{207, 207, 207}
\definecolor{ipython_bg}{RGB}{247, 247, 247}
\definecolor{ipython_red}{RGB}{186, 33, 33}
\definecolor{ipython_green}{RGB}{0, 128, 0}
\definecolor{ipython_cyan}{RGB}{64, 128, 128}
\definecolor{ipython_purple}{RGB}{170, 34, 255}

\newcommand{\jeffgrumbler}[2]{\textcolor{blue}{\bf #1: #2}}
\newcommand{\jeff}[1]{\jeffgrumbler{Jeff}{#1}}

\renewcommand{\jeffgrumbler}[2]{}
\renewcommand{\jeff}[1]{}

\newcommand{\nic}[1]{\textcolor{black}{#1}}
\newcommand{\craig}[1]{\textcolor{black}{#1}}
\newcommand{\craigcrc}[1]{\textcolor{black}{#1}}
\newcommand{\crc}[1]{\textcolor{black}{#1}}

\newcommand{\nicnew}[1]{\textcolor{black}{#1}}

\newcommand{\retrieve}{\textsf{Retrieve}}
\newcommand{\rewrite}{\textsf{Rewrite}}
\newcommand{\expand}{\textsf{Expand}}
\newcommand{\extract}{\textsf{Extract}}
\newcommand{\rerank}{\textsf{Rerank}}

\newcommand{\pipe}{\textsf{Pipe}}

\newcommand{\fit}{\textsf{fit}}
\newcommand{\experiment}{\textsf{Experiment}}

\usepackage{xcolor,colortbl}

\usepackage{cleveref}

\AtBeginDocument{\DeclareCaptionSubType{lstlisting}}
\crefname{sublstlisting}{listing}{listings}
\Crefname{sublstlisting}{Listing}{Listings}

\usepackage[show]{chato-notes}

\copyrightyear{2020}
\acmYear{2020}
\setcopyright{acmcopyright}
\acmConference[ICTIR '20] {2020 ACM SIGIR International Conference on the Theory of Information Retrieval}{September 14--17, 2020}{Virtual Event, Norway}
\acmBooktitle{2020 ACM SIGIR International Conference on the Theory of Information Retrieval (ICTIR '20), September 14--17, 2020, Virtual Event, Norway}
\acmDOI{10.1145/3409256.3409829}
\acmISBN{978-1-4503-8067-6/20/09}

\usepackage{subcaption}

\usepackage{marginnote}
\newcommand{\pageenlarge}[1]{\enlargethispage{#1\baselineskip}}

\usepackage{xspace}

\def\terrier{Terrier\xspace}
\def\name{PyTerrier\xspace}

\begin{document}
\title{Declarative Experimentation in \\ Information Retrieval using \name}
\fancyhead{}

\author{Craig Macdonald}
\affiliation{%
    \institution{University of Glasgow}
}

\author{Nicola Tonellotto}
\affiliation{%
    \institution{University of Pisa}
}

\begin{abstract}
\looseness -1 The advent of deep machine learning platforms such as Tensorflow and Pytorch, developed in expressive high-level languages such as Python, have allowed more expressive representations of deep neural network architectures. We argue that such a \nicnew{powerful} formalism is missing in information retrieval \nicnew{(IR)}, and propose a framework called \craigcrc{\name that allows} advanced retrieval pipelines to be expressed, and evaluated, in a declarative manner close to their conceptual design. Like the \nicnew{aforementioned frameworks} that compile deep learning experiments into primitive GPU operations, our framework targets \nicnew{IR} platforms as backends in order to execute and evaluate retrieval pipelines. Further, we can automatically optimise the retrieval pipelines to increase their efficiency to suite a particular IR platform backend. Our experiments, conducted on TREC Robust and ClueWeb09 test collections, demonstrate the efficiency benefits of these optimisations for retrieval pipelines involving both the Anserini and Terrier IR platforms.
\end{abstract}
\maketitle


\section{Introduction}\pageenlarge{2}
Information retrieval (IR) is classically an empirical science. Offline experiments towards enhancing retrieval effectiveness have been easily made possible through use of test collections with documents judged for relevance by human assessors, as typified by the TREC, CLEF, NCTIR evaluation forums.

On the other hand, machine learning has experienced even greater growth, with applications to many areas of science, driven by the availability of good datasets~\cite{tensorflow}, as well as platforms that allow easy development and application of machine learned models. In recent years, there has been a focus on the development and application of deep learning frameworks written in high-level languages, including Lua (Torch), but particularly Python (Tensorflow and Pytorch). Using \crc{such} expressive high-level languages allow complex deep neural network architectures with various matrix operations to be expressed using familiar programming paradigms, for instance, adding matrices using a \texttt{+} operator, or adding several hidden layers using a \texttt{for} loop to add objects to a list.

\begin{figure}
\centering
\includegraphics[scale=0.7]{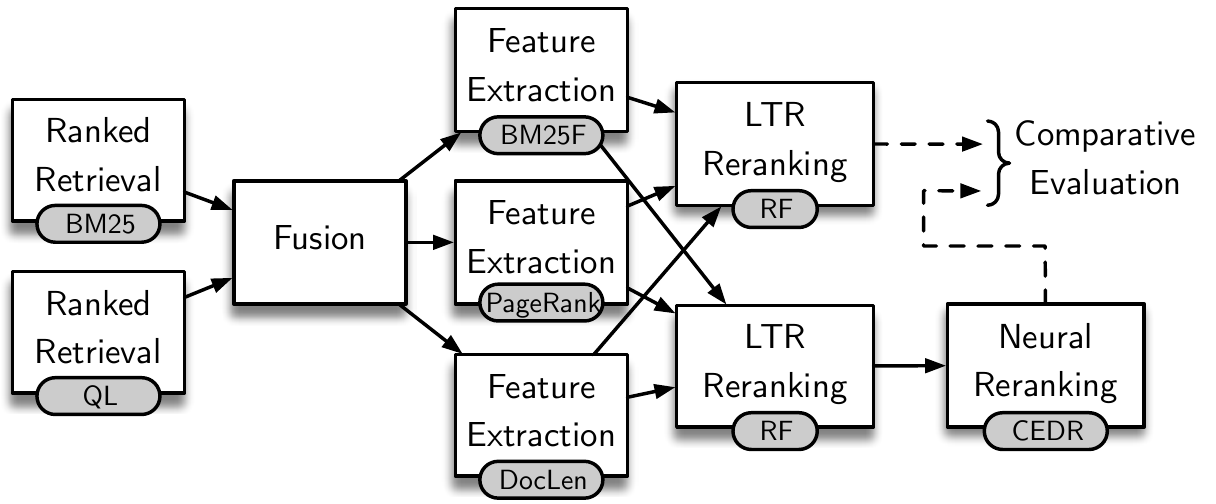}
\vspace{-2\baselineskip}
\caption{An experiment comparing two complex retrieval pipelines, with learning-to-rank and neural re-ranking.}\label{fig:dag}\vspace{-1\baselineskip}
\end{figure}

\looseness -1 We argue that adoption of such an expressive high-level languages are missing from many of the available IR platforms, and hence we are unable to perform wide-ranging experiments with the ease of our machine learning compatriots. End-to-end retrieval evaluation is not easy to perform.
\nic{For example, Figure~\ref{fig:dag} depicts, in the form of a directed acyclic graph (DAG), an example IR experimental workflow to compare the effectiveness of two IR systems involving the setup and execution of different classical IR techniques, including ranked retrieval with different weighting models, fusion, features extraction, learning to rank (LTR) algorithms and neural re-ranking. Typically, conducting such experiments involves editing several configuration files, running various commands to generate result files to be fed to the following stages, and eventually invoking {\tt trec\_eval} to evaluate the experimental outcomes.}
Configuration is spread across several files making reproducibility difficult.

Yet reproducibility is key to impactful science. Ferro \& Kelly~\cite{ferro} define reproducibility as the ability for a different team to reproduce the measurement in a different experimental setup. Therefore, focussing evaluation solely on datasets that extract key aspects of a problem using a standard dataset -- for instance, evaluating LTR techniques solely on LETOR  datasets~\cite{DBLP:journals/corr/QinL13} with common features -- does not allow us to understand the wider context, such as how an approach will fare when integrated into a fully-fledged search engine's retrieval stack. This highlights the importance of end-to-end retrieval experiments -- understanding what data are needed for a given approach, and how it interacts with others components (e.g., how many documents should be re-ranked~\cite{whens} for a LTR approach), reduces the uncertainties when a technique should be deployed to an operational search engine.

\looseness -1 \pageenlarge{2} Thus, we argue that a succinct manner of describing a retrieval \nic{experiment}, in a {\em conceptual} yet familiar way, should allow more IR researchers with increased ability to develop techniques that can be easily integrated, and evaluated, in an end-to-end fashion. Hence, in this paper, we describe a new framework called \name\footnote{\craigcrc{\url{https://github.com/terrier-org/pyterrier}}} for expressing \nic{IR experiments with composable \emph{pipelines}}. Similar to Tensorflow and Pytorch, it uses Python as a high-level language for operationalising of experiments. Moreover, we use standard operators to combine objects representing retrieval building blocks called {\em transformers}, allowing advanced retrieval pipelines to be specified in a declarative rather than procedural fashion. We have initially instantiated \name on two existing IR platforms (Anserini and \terrier), and demonstrate that the transformers and operators can be used to retrieve from, and even combine, multiple systems.

Moreover, the expression of a retrieval pipeline in a high-level language using transformers and operators, forms a DAG of basic IR operations (retrieve, re-rank, combine results, rewrite query, etc). This DAG can be rewritten, optimised and/or passed to the underlying search engine. Hence, a complex retrieval architecture trialled in \name might be compiled down to \nic{an optimized configuration}
of the underlying search engine(s).  Indeed, we show how simple optimisations steps on the DAG result in markedly enhanced efficiency \nic{when such system-specific optimisations are employed.}


\nic{Therefore, this paper contributes a conceptual framework for a different way of envisaging and executing information retrieval experiments. We \crc{propose a clear semantics} for different IR experiment's building blocks and their composition. The proposed framework is easily extensible, since it allows the definition and inclusion of new IR activities.}
\nic{Moreover, we provide an implementation of our proposed framework in Python, supporting the compilation of IR pipelines on top of two widely-used Java-based IR platforms to conduct fast retrieval operations.}
Finally, we show how these operations can be optimised --  while retaining their semantics -- by applying {\em graph rewriting patterns} targetting the underlying IR platform operations. Moreover, it allows easy integration of deep learning techniques into the retrieval pipeline, thereby allowing to enhance the reproducibilty of retrieval technologies.

\looseness -1 The remainder of this paper is structured as follows: Section~\ref{sec:related} positions our framework with respect to different standard toolkits, for data science, big data, machine learning and information retrieval; In Section~\ref{sec:tr_ops}, we formally describe our framework in terms of transformers, operators, and experiments. Section~\ref{sec:rewriting} \nic{discusses how pipelines can be compiled and optimised through \emph{graph rewriting} to be efficiently implemented using less calls to the underlying IR platforms.}
Section~\ref{sec:experiments} demonstrates the empirical efficiency in these pipeline rewrites through experiments using \name based on two underlying search engines, namely Anserini and Terrier. We provide concluding remarks and detail future improvements in Section~\ref{sec:conclusions}.


\pageenlarge{2}\section{Related Work}\label{sec:related}

\looseness -1 IR platforms have a long history, dating back to at least SMART~\cite{smart}. These days, among the open source platforms, Apache Lucene is widely deployed. Implemented in Java, it provides indexing and single-search APIs, and in recent years has adopted BM25 along with LTR~\cite{luceneltr} and dynamic pruning techniques~\cite{DBLP:conf/ecir/GrandMFL20}. However, its ability to handle standard test collections was for many years a known limitation~\cite{lucene4ir}, and has been advanced by efforts such as Anserini~\cite{anserini2017}. Indeed Anserini facilitates the deployment of a number of replicable information retrieval techniques, on standard test collection, on top of the Lucene backend.

Among the other commonly used academic platform are Lemur/Indri (implemented \crc{in} C) -- along  the closely related Galago (implemented in Java)~\cite{cartright2012galago} -- as well as Terrier (implemented in Java)~\cite{macdonald12terrier} and PISA~\cite{MSMS2019} (implemented in C++). We note that while Python ``bindings'' exist for various platforms, including Indri, Galago and Anserini, there are no serious contenders for IR platforms written natively in Python. We believe there are several reasons for this, including the challenge of indexing and retrieving from corpora that contains 1-10 million documents, as well as the more recent maturity of Python. Indri and its modern replacement Galago also have rich (domain-specific) query languages that allow the expression of complex retrieval operations.

\nic{All \crc{of} the discussed IR platforms mix the design of experimental retrieval activities with the implementation and optimisations required to make such activities efficient. This approach has been shown to limit the reproducibility of IR experiments. For example, M\"uhlisen \emph{et al.}~\cite{10.1145/2600428.2609460} show that different implementations of the same BM25 weighting models in different IR platforms result in different values for the same effectiveness metric.}
\nic{They propose to decouple the IR experiments from the IR platform implementation by storing the inverted index in a column-oriented relational database and by implementing ranking models using SQL. Kamphuis \& de Vries~\cite{ggraph} take a step forward and propose the adoption of an IR-specific declarative language to provide higher level abstractions in the implementation of the IR experiments based on a graph query language. In contrast to this and the Indri/Galago domain-specific \crc{query} language, we propose a declarative framework to express basic retrieval operations and their composition using queries and documents as inputs and outputs. It is built upon Python, which is expressive, readily accessible and allows integration with other modern Python toolkits such as those for deep learning. Together, this allows for rapid prototyping and improved reproducibility in IR. We then show how the elements of the proposed declarative language can be compiled into a DAG representation, which can be efficiently implemented on specific IR platforms.}

\looseness -1 More generally, Python is popular among other branches of data science and machine learning \crc{--} it has no need to compile, and hence allows researchers/developers to easily adjust their code and re-run, in an agile manner, often using notebook environments such as Jupyter or Google Colab; Standard toolkits such as Pandas (for structured data processing in relational dataframes) and scikit-learn (for machine learning) exemplify this approach to data science.


\looseness -1 To allow the efficient application of data processing and machine learning at scale, Apache Spark overcame some of the disadvantages of the store-then-compute MapReduce programming paradigm; Apache Spark, which has bindings in Java, Scala and Python, allows structured data processing operations to be vastly parallelised across a cluster of machines~\cite{10.5555/1863103.1863113}. As a functional language, expressions in Apache Spark, even in Python, are compiled into an execution plan, which is then distributed to different compute machines. We are inspired by this notion of an execution plan.
\jeff{What about SQL, PIG, Flume, and Beam? For data processing the trend has been away from declarative back towards high-level procedural (like Spark/Beam) because of their flexibility.}

\pageenlarge{2} \looseness -1 In recent times, Tensorflow~\cite{tensorflow} and PyTorch have been the dominant frameworks for neural machine learning. Both are based on primitive operations (add, multiple, concatenate) on tensors, which are expressed in Python by overloading the math operators in the language for the tensor objects; higher level tensor operations such as recurrent \crc{units}, attention etc., can be achieved using higher level objects. Thus a domain-specific programming environment is created in Python. The object graph instantiated by the tensor objects and operators creates a dataflow \crc{DAG}, which can be compiled into GPU operations  for efficient computation.

\looseness -1 In this work, we are inspired by these existing deep learning frameworks, as we both instantiate a domain-specific programming environment in Python, but this time suited for \crc{IR} experiments. Moreover, similar to deep learning, we observe that search pipelines expressed in this manner \crc{form DAGs, which} can be compiled into more efficient low-level search engine operations.


\looseness -1 The most similar work to our own is that in Terrier-Spark~\cite{terrierspark1, terrierspark2}, where retrieval pipelines for the Terrier platform were created in Scala using Apache Spark. In that work, retrieval operation were expressed as operations on dataframes (relations). However, adoption of that framework was hindered by two factors: firstly, the use of Apache Spark, which is designed for processing massive scale datasets, and introduces significant interactive overheads making Terrier-Spark unsuitable for notebook-style agile development; secondly, the use of the Scala programming language, which is not as popular as Python. In this paper, we \craigcrc{extend} the notion of retrieval operations on dataframes, but instead, operate on Python, \craigcrc{and create} a domain-specific programming environment where complex retrieval pipelines can be formulated as operations on Python objects.


\pageenlarge{2}\section{Declarative Retrieval Operations}\label{sec:tr_ops}

\nic{In this section we discuss the IR data model we use to represent queries and documents (Sec.~\ref{ssec:irmodel}), we introduce IR \emph{transformers} to express IR operation as the basic element of our declarative IR language (Sec.~\ref{ssec:transformers}) and \emph{operators} to compose transformers into more complex IR \emph{pipelines} (Sec.~\ref{ssec:operators}), and finally we discuss how to design IR experiments with our declarative IR language (Sec.~\ref{ssec:experiments}).}

\subsection{IR Data Model}\label{ssec:irmodel}


The basic elements of an IR systems are \emph{queries} and \emph{documents}. A query $q$ is a textual representation of an information need, while a document $d$ is a textual source of information.
From an object relational model perspective, $q$ and $d$ are \emph{tuples}, i.e., finite sequences of attributes, whose domains and values depends on their associated metadata. These metadata can vary, but in the following we will assume a query $q$ encapsulates at least the query text $q.text$ and the query identifier $q.id$, which also \crc{constitutes} the primary key.
Similarly, we will assume a document $d$ encapsulates at least its full text $d.text$, and the document identifier $d.id$, which also represents the primary key. Both queries and documents may have additional attributes, such as the number of unique terms, the document's \crc{URL, title and TREC docno, title,} and so on.
We denote with $Q = [q_1, q_2, \ldots]$ a finite ordered list of queries, and with $D = [d_1, d_2, \ldots]$ a finite ordered list of documents. From an object relational model perspective, both $Q$ and $D$ are \emph{relations}.


\looseness -1 Further, to allow evaluation, we formally define relevance assessments (in TREC parlance, \emph{qrels}). A relevance assessment $ra$ is a tuple whose attributes include a query id, a document id (together forming the primary key) and a relevance label. A finite set of relevance assessments is denoted as $RA = \{ra_1, ra_2, \ldots\}$. Finally, given a list of queries $Q$, we denote with $R = [r_1, r_2, \ldots]$ the list of \emph{ranked results} for the queries. Each retrieved result $r$ associates a query $r.q$ and a document $r.d$ with a relevance score $r.s$ that defines the ranking order. The primary key of a retrieved results list is the pair $(q.id, d.id)$. Moreover, a retrieved results list \craigcrc{may exhibit} additional metadata such as the query-document \emph{features} commonly exploited in LTR scenarios~\cite{Liu09ftir}.





\subsection{IR Transformers}\label{ssec:transformers}
We exploit the IR data model discussed above to express retrieval operations in an IR system as transformations between queries and document. To this end, we leverage \emph{function objects}. A function object is a construct allowing an object to be invoked or called as if it were an ordinary function. A function object
has properties depending on the function object's specific implementation. 
\nicnew{These properties allow the explicit declaration of the configuration parameters of each instantiated function object.}

We build our declarative IR language on a generic function object we call \crc{a} \emph{transformer}. In general, a transformer $f: Q \times R \mapsto Q \times R$ takes as input a list of queries $Q$ and a list of retrieved documents $R$, and returns another list of queries $Q'$ and another list of documents $R'$.
Depending on the specific implementation of a transformer, both inputs and output can be partially specified, i.e., input can be just $Q$ or $R$ and/or output can be just $Q'$ or $R'$.
An optional input can be omitted, i.e., the input is assumed to be an empty list, or, if present, it is ignored by the transformer. An optional output is ignored by the transformer, and it is a verbatim copy of the corresponding input.
As we will see, this definition of transformer is general enough to allow the composition of transformers.

In the following, we describe some classes of transformers to show how commonly deployed operations in IR experiments can be implemented as transformers.

\paragraph{Basic retrieval} A classical search operation can be expressed as a function $\retrieve: Q \mapsto R$:
\begin{align}
    & R' = \retrieve().\textsf{transform}(Q).
\end{align}
In doing so, the \retrieve~function transforms a list of queries $Q$ in a list of retrieved results $R$, i.e., a set of documents $D$ for each query $q$ in $Q$. We may instantiate the \retrieve~operation in many ways, for instance using different weighting models (selecting e.g., BM25, TF.IDF, language modelling \craigcrc{and their parameters} as properties of the transformer) or even using different retrieval systems (Indri, Terrier, Anserini, etc.). 
\begin{equation}
    R' = \retrieve(\textsf{"BM25"}).\textsf{transform}(Q). \label{eqn:RetBM25}
\end{equation}
Moreover, we assume that \textsf{transform}() is the default method in our transformer objects, and hence need not be specified, i.e., Equation~\eqref{eqn:RetBM25} is equivalent to:
\begin{align}
R' = \retrieve(\textsf{"BM25"})(Q)
\end{align}

Note that, in the following, if no properties are specified, we suppress the empty $\textsf{()}$ to improve readability.

\pageenlarge{2}\paragraph{Query Rewriting}

In many cases, a query might be rewritten by the IR system before passing to the ranking component. For instance, the sequential dependence proximity model~\cite{sdm} adds operators such as the Indri {\tt\#1} and {\tt \#uw8} operators containing pairs and sequences of query terms, in order to boost the scores of documents where the query terms appear in close proximity. Similarly, Peng \emph{et al}.~\cite{csse} describe a query rewriting operation, known as context-sensitive stemming, where for some query terms, alternative inflections are added to the query.
Rewriting can easily be expressed \crc{as a transformer like} $\rewrite: Q \mapsto Q$, e.g.,:
\begin{equation}
Q' = \rewrite(Q).
\end{equation}

\paragraph{Query Expansion}
In pseudo-relevance feedback (PRF) query expansion, additional query terms can be added to the query based on how they occur in documents highly-ranked for the initial query. The identification of the refined query, can be expressed as \crc{a} transformer, $\expand: Q \times R \mapsto Q$, e.g.:
\begin{equation}
Q' = \expand(Q,R).
\end{equation}
The reweighted query \crc{should then be} re-executed on the original index. Hence, the whole PRF process can be expressed as a combination of three transformers: a first \retrieve~transformer, a \expand~transformer, which takes the queries $Q$ and retrieved documents $R$, and calculates reformulated queries $Q'$ by examination of the top-ranked documents for each query, and a second \retrieve~transformer, to process the reformulated queries, e.g.,
\begin{equation}\label{eq:prf}
R' = \retrieve(\expand(\retrieve(Q))).
\end{equation}

\paragraph{Feature extraction}

With the advent of LTR, multi-stage ranking pipelines have become commonplace in IR experiments. In the classical LTR paradigm, for a given query, $k$ documents are ranked by an initial retrieval approach to form a {\em candidate set}; $k=1000$ and BM25 form a typical setup~\cite{whens}. Upon this candidate set, a number of additional retrieval features are extracted or calculated. For instance, PageRank or URL length are examples of {\em query independent} features that might be used in web search settings; proximity, field-based weighting models~\cite{macdonald2013tois} are examples {\em query dependent} features.
Both query-independent and query-dependent feature extraction can be expressed as transformers. Without loss of generality, we can use a single transformer $\extract: Q \times R \mapsto Q \times R$ to encompass all feature extraction processes, e.g.,
\begin{equation}
Q', R' = \extract(Q, R),
\end{equation}
where $Q$ is optional when extracting query-independent features.

\paragraph{Reranking}

In multi-stage ranking pipelines, after feature extraction a candidate set of documents is re-ranked to boost effectiveness. Beside reranking using a LTR technique such as LambdaMART, more recently neural re-rankers such as BERT (e.g.,~\cite{CEDR2019}) are being increasingly widely used for improved effectiveness.
In all cases, a re-ranker takes an input set of documents, and computes a new score for that set of documents, and hence a reranker can be expressed as the transformer $\rerank: Q \times R \mapsto R$, and a two-stage retrieval pipeline with feature extraction can be expressed as a combination of transformers, e.g.,
\begin{equation}\label{eq:rerank}
R' = \rerank(\extract(\retrieve(Q))).
\end{equation}

Most re-rankers exploit machine learning techniques, hence they must be trained on some test data. To trigger the training of a re-ranker, this transformer \craigcrc{exposes} a method to estimate the model parameters to be used in subsequent IR experiments:
\begin{equation}
    \rerank.\fit(Q_{train}, RA_{train}, Q_{valid}, RA_{valid}), \label{eq:fit}
\end{equation}
where the parameters $Q_{train}, RA_{train}, Q_{valid}$, and $RA_{valid}$ denote the training queries and qrels, and the validation queries and qrels, respectively, used to train the underlying machine-learned model.

Table~\ref{tab:transformers_nic} summarises the transformer classes presented. The optional input/output queries/retrieved documents are in parenthesis. \craigcrc{Finally, we note that any arbitrary function that takes $Q$ \crc{and/or} $R$ and returns $Q$ \craig{and/or} $R$ can be used as a transformer, thereby allowing easy extensibility.}

\begin{table}[tb!]
\centering
\caption{Classes of Transformers. The optional input/output queries/retrieved documents are in parentheses.} \vspace{-\baselineskip}
\label{tab:transformers_nic}
\begin{tabular}{ccl}
Input & Output & Transformer \\
\toprule
$Q~(\times~R)$ & $Q'~(\times~R)$ & Query rewriting \\
$Q~(\times~R)$ & $(Q~\times)~R'$ & Basic Retrieval \\
$Q~\times~R$ & $Q'~(\times~R)$ & Query expansion \\
$Q~\times~R$ & $(Q~\times)~R'$ & Re-ranking \\
$Q~\times~R$ & $Q' \times R'$ & Feature extraction \\
\bottomrule
\end{tabular}
\end{table}


\pageenlarge{2}\subsection{Operators for Transformers}\label{ssec:operators}

The notation for nested transformers calls in Equation~\eqref{eq:rerank} is difficult to write and hides the fact that a first-stage retrieval occurs before a second-stage re-ranking. To make it easy to combine different transformers in a succinct and easily understandable manner, we are inspired by deep learning frameworks towards creating succinct {\em pipelines} of IR transformations by \emph{operator overloading}. In this way, we can use Python-like operators to allow simple notations for retrieval pipelines.
In the following, we leverage some notations from relational algebra to describe these operators, as follows:
\begin{itemize}
    \item Let $R_1 \bowtie R_2$ denote the natural join between two retrieved results lists. The result of the natural join is the set of all combinations of retrieved results in $R_1$ and $R_2$ that are equal on both of their composite $(q.id, d.id)$ primary key attributes;
    \item \looseness -1 \nic{Let $_a\Gamma_{b}(R)$ denote the sorting of tuples in $R$ according to the ascending values of attribute $b$ after grouping by attribute $a$.} 
    \item \nic{Let $_a\sigma_K(R)$ denote the selection of the first $K$ tuples in $R$ after grouping by attribute $a$. } 
    \item Let $R[f(a_1, ..) \to b]$ denote the transformation of one or more attributes $a_1, ..$ of the tuples in $R$ to a new attribute $b$ according to function $f(\cdot)$.
    \item Following \cite[p.236]{Silberschatz}, let $_{a}\mathcal{G}_{\text{op}(b)}(R)$ denote the application of operator $\text{op}(\cdot)$ to attribute $b$ of the tuples in $R$ after grouping by attribute $a$, i.e., equivalent to a SQL statement projecting an aggregate function on $b$ after a GROUP~BY~on~$a$.
\end{itemize}
\pageenlarge{2} We now describe the transformer operators we have defined. A summary is provided in Table~\ref{tab:operators_nic}.

\begin{table}[tb!]
    \centering
     \caption{\name operators for combining transformers.}    \label{tab:operators_nic} \vspace{-\baselineskip}
    \begin{tabular}{ccp{5cm}}
    \toprule
     Op.\ & Name & Description \\
    \midrule
        \texttt{>}\texttt{>} & \textit{then} & Pass the output from one transformer to the next transformer\\
        \texttt{+} & \textit{linear combine} & Sum the query-document scores of the two retrieved results lists \\
        \texttt{*} & \textit{scalar product} & Multiply the query-document scores of a retrieved results list by a scalar \\
        \texttt{**} & \textit{feature union} & Combine two retrieved results lists as features \\
        \texttt{|} & \textit{set union} & Make the set union of documents from the two retrieved results lists \\
        \texttt{\&} & \textit{set intersection} & Make the set intersection of the two retrieved results lists \\
        \texttt{\%} & \textit{rank cutoff} & Shorten a retrieved results list to the first $K$ elements \\
        \texttt{\string^} & \textit{concatenate} & Add the retrieved results list from one transformer to the bottom of the other \\
    \bottomrule
    \end{tabular}\vspace{-\baselineskip}
\end{table}

\emph{Linear combination} (\texttt{+}). This operator allows two retrieval sets to be linearly combined, to support CombSUM data fusion or linear interpolation of relevance scores. If $T_1$ and $T_2$ represent transformers returning the same set of queries $Q$ as in input, then
    \begin{align*}
        Q, R' &= (T_1 + T_2)(Q,R) :=\\
        & R_1 = T_1(Q,R), \quad R_2 = T_2(Q,R)\\
        & R' = (R_1 \bowtie R_2)[s_1+s_2 \to s]
    \end{align*}

\emph{Scalar product} (\texttt{*}). This operator allows the scores of a retrieval set to be multiplied by a scalar value $\alpha$:
    \begin{align*}
        Q', R' &= (\alpha~\texttt{*}~T)(Q,R) :=\\
        & Q', R_1 = T(Q,R) \\
        & R' = R_1[\alpha s \to s]
    \end{align*}
This permits the weighting of systems within a linear combination.

\emph{Set union (\texttt{|}) and intersection (\texttt{\&})}. These operators allow two retrieval sets to be combined -- for example, combining the \crc{documents obtained} with and without PRF for use as a candidate set~\cite{dietz_ictir2019}. Due to their inherent set properties, the resulting document scores are not defined, e.g., the scores assume the special value $\perp$. Thus, these operators are intended for use \crc{with} further (re-)ranking. More specifically, if $T_1$ and $T_2$ represent transformers returning the same set of queries $Q$ as in input, then the set union would be defined as:
    \begin{align*}
        Q, R' &= (T_1~\texttt{|}~T_2)(Q,R) :=\\
        & R_1 = T_1(Q,R), \quad R_2 = T_2(Q,R)\\
        & R' = (R_1 \cup R_2)[\perp \to s]
    \end{align*}

\emph{Rank cutoff} (\texttt{\%}). This operator allows a ranking to be truncated at a given rank $K$. Given a retrieval results list produced by the transformer $T$, the result of the $T \texttt{ \% } K$ operation is a retrieval result containing, for each query, the $K$ documents returned by $T$ with the highest scores:
    \begin{align*}
        Q, R' &= (Q,R) \texttt{ \% } K :=\\
        & R_1 = _{q.id}\Gamma_{-s}(R)\\
        & R' =  _{q.id}\sigma_K(R_1).
    \end{align*}

\pageenlarge{2}\emph{Concatenate} (\texttt{\string^}). This operator appends a second ranking after a first one for each query. This is useful in cases where, for instance, we may re-rank a few documents using an expensive approach, such as BERT, then append the remainder of the ranking from the baseline. Documents appearing in the first ranking for each query are removed from the second ranking. Documents from the second ranking have their scores adjusted such that the \crc{highest ranked remaining document} from the second ranking has a score just less than the lowest ranked document from the first ranking. More formally:
    \begin{align*}
        Q, R' &= (T_1 \texttt{ \string^ } T_2)(Q,R) :=\\
        & R_1 = T_1(Q,R), \quad R_2 = T_2(Q,R)\\
        & \underline{S} = _{q.id}\mathcal{G}_{\min(s)}(R_1), \quad \overline{S} = _{q.id}\mathcal{G}_{\max(s)}(R_2 - R_1) \\
        & R' = R_1 \cup \big((R_2 - R_1) \bowtie \underline{S} \bowtie \overline{S}\big) [r_2.s - \underline{s}.s + \overline{s}.s - \epsilon \to r_2.s]
    \end{align*}
\looseness -1 where
$\epsilon$ is a small constant, e.g., $\epsilon=0.001$, used to represent the minimum score difference between documents coming from $R_1$ and $R_2$. 

\emph{Feature union} (\texttt{**}). This operator is intended to allow different retrieval systems to be composed as features for LTR.  More specifically, if $T_1$ and $T_2$ represent transformers returning the same set of queries $Q$ \crc{and documents $R$ as is} input, then
    \begin{align*}
        Q, R' &= (T_1 \texttt{ ** } T_2)(Q,R) :=\\
        & R_1 = T_1(Q,R), \quad R_2 = T_2(Q,R)\\
        & R' = (R_1 \bowtie R_2)[ [f_1,f_2] \to f]
    \end{align*}
The resulting retrieved result list combines, for each $(q.id, d.id)$ tuple, the features $f_1$ and $f_2$, e.g., the metadata, of the two input retrieved result lists into a new list of features $f$.


\emph{Composition} (\texttt{>}\texttt{>}). This operator denotes that the output of an IR transformer should be used as input \crc{to} another IR transformer. More specifically, if $T_1$ and $T_2$ represent transformers, then
    \begin{align*}
        Q', R' &= (T_1 >> T_2)(Q, R) := \\
        &Q_1, R_1 = T_1(Q,R) \\
        &Q', R' = T_2(Q_1,R_1).
    \end{align*}
For brevity, we also describe the composition operator as \emph{then}.
This operator allows to ``chain'' transformers and/or operator together, to define experimental \emph{pipelines}. As an illustration, use of the composition operator, allows Equation~\eqref{eq:rerank} to be succinctly written as:
    \begin{align*}
        & \pipe = \retrieve >> \rerank\\
        & R' = \pipe(Q)
    \end{align*}
Note that the type of \pipe{} is also a transformer, and hence all operators can be applied upon that transformer. \nic{Moreover, if at least one of the composed transformers expose a \fit{} method as in Eq.~\eqref{eq:fit}, then the composed pipeline exposes a \fit{} method as well. This method triggers the training of all the machine-learned \rerank{} transformers in the pipeline. Other transformers are applied as necessary, in order to make the appropriate transformation of the queries into the required inputs for the \fit{} method.}

\pageenlarge{2}\subsection{\craigcrc{Experiment Abstraction} on Pipelines}\label{ssec:experiments}

\looseness -1 Having defined the suite of transformers and operators available, we now turn to actually running IR retrieval experiments. In a procedural fashion, a retrieval pipeline can be evaluated in three steps:
\begin{itemize}
    \item obtain the queries $Q$ and corresponding relevance the assessments $RA$;
    \item transform those queries into results using the pipeline, let say $R = \pipe(Q)$;
    \item apply an evaluation tool, such as the ubiquitous \texttt{trec\_eval} tools -- or its Python bindings \texttt{pytrec\_eval}~\cite{VanGysel2018pytreceval} -- on $RA$ and $R$, to obtain effectiveness measures such as MAP or NDCG.
\end{itemize}

Application of these three procedural steps might be considered laborious. \nic{Besides a sound knowledge of the specific IR system software to used, experiments are typically run using multiple scripting tools generating a lot of output files to be evaluated and compared to the outcomes of different experiments with similar but different implementations.} Instead, we are inspired by the Cornac framework\footnote{\url{https://cornac.preferred.ai/}} for conducting comparative recommender system experiments. Cornac provides a succinct \experiment{} \nic{abstraction} that allows many recommender systems to be evaluated using the same datasets for the same evaluation measures, while ensuring fair setup across matter such as cross-validation splits.

To this end, we define an \experiment{} function in \name{}, which can apply a list of retrieval pipelines upon a common set of queries, and evaluates the resulting result set from each pipeline to obtain a common set of evaluation measures. Our \experiment{} function builds upon the \texttt{pytrec\_eval}~\cite{VanGysel2018pytreceval} tool. It is also of note that \experiment{} acts as a trigger for the application of the pipelines upon a query set. \nic{The syntax for our proposed implementation of the \experiment~abstraction is:}
\begin{align*}
    \experiment([\pipe_1, \pipe_2], Q, RA, [\textsf{"map"}, \textsf{"ndcg"}]).
\end{align*}
The output of an \experiment{} execution is a table comparing the specified retrieval pipelines side-by-side.
\begin{lstlisting}[language=Python,float=*,numbers=left,caption={Example experiment for the document ranking task of the TREC 2019 Deep Learning track.},label=lst:one]
first_pass = Retrieve(index, "BM25")                        # initial retrieval
# prf defines the candidate documents to re-rank using the additional features
prf = first_pass >> RM3(index) >> Retrieve(index, "BM25")
sdm = SequentialDependence >> Retrieve(index, "BM25")       # rewrites the query to use proximity operators
bert = CEDRPipeline("vanilla_bert")                         # applies a BERT re-ranker
ltr = xgBoostPipeline(xgBoost({"rank": "ndcg"}))            # defines and configures the LambdaMART LTR stage
# combine the full pipeline, using query expansion scores, proximity and BERT as features.
full_pipeline = prf >> (sdm **  bert) >> ltr
full_pipeline.fit(tr_topics, tr_qrels, va_topics, va_qrels) # train the pipeline
# evaluate the pipelines. Report MAP and NDCG@10
Experiment([first_pass, prf, full_pipeline], test_topics, test_qrels, metrics=["ndcg_cut_10", "map"])
\end{lstlisting}

While an \experiment~does not currently handle the fitting (training) of the pipeline, it is easy to consider variants that automatically handle $k$-fold cross validation. A further variant might be the introduction of a grid search functionality to determine the best settings for different components in the pipeline. Due to the compositional nature of a retrieval pipeline, the grid search would be able to cache the outcomes of earlier stages in the pipeline, such that later retrieval components could be varied without re-execution of all pipeline stages.

\pageenlarge{0} Combined with the previously described transformers and operators, our proposed \experiment{} function allows the researcher to focus on {\em what} is being evaluated, i.e., the stages of the retrieval pipeline, rather than on the order of execution. Indeed, through the resulting domain-specific programming environment, researchers can design IR pipelines in terms of the concepts of the approach (combine these models to make a good candidate set; express these models as re-rankers; combine and learn a LTR model using these re-rankers as features). Each component can be simply implemented as a transformer operating on queries and documents.

In essence, we believe that the aforementioned types of transformers and operators allows to address a plethora of different IR operations. We define primitive search operations, such as search, rewrite, rerank, which \crc{can} be easily implemented using standard search engine toolkits - indeed, we already have implementations for the Anserini and Terrier platforms. Moreover, instantiation of a learned model can be easily achieved by appending final transformers for learned methods for sci-kit Learn, (e.g., Random Forests) or xgBoost~\cite{xgboost} (e.g., the LambdaMART LTR algorithm~\cite{burges2010from}). We have also implemented transformer objects for neural re-ranking implementations such as CEDR~\cite{CEDR2019}.

\looseness -1 Listing~\ref{lst:one} provides an example instantiation of a retrieval experiment, demonstrating how different retrieval transformers might be combined into a comprehensive, yet easily understandable, retrieval pipeline. This particular example, which might be applied to the TREC Deep Learning Track, demonstrates a composed pipeline involving pseudo-relevance feedback, BERT and LambdaMART.


However, the capabilities of different search engine toolkits differ, and hence \craigcrc{there} may be more efficient ways to instantiate the same pipeline. In the next section,
\nic{we further elaborate on this topic,} 
\nic{demonstrating how different toolkits can be used and optimised to efficiently implement the same pipeline.} 
This allows the conceptual design of a retrieval pipeline -- as expressed using transformers and operators -- to diverge from its logical implementation, as executed upon the IR toolkits.

\section{Implementing Transformers and Operators}\label{sec:rewriting}

\nic{Our domain-specific declarative environment for IR search experiments allows us to focus on the logical design of our experiments. The output of this design activity is a computational data-flow graph, with search operations as nodes, data dependencies between search operations as edges, where queries and documents are passed along edges. Some of these operations represent transformers for primitive search operations such as search, rerank, rewrite (as summarised in Table~\ref{tab:transformers_nic}), while others represent operators to combine these transformers in different ways (as summarised in Table~\ref{tab:operators_nic}). } 

\pageenlarge{0} \nic{For the actual execution of IR retrieval experiments, the logical design of the experiments, i.e., the composition of pipelines, must be is \emph{compiled} targeting a specific IR software platforms. Depending on the execution platforms, transformer and operators can be combined together to allow more efficient execution of the experiment. }
\craig{To implement the notions of pipeline compilation and optimisation, we use a pattern matching algorithm to identify patterns of pipeline expressions on a given {\em subject} pipeline, and apply \emph{graph rewriting patterns} to create the optimised version of the subject pipeline for a given IR platform. Each pattern applies equivalence rules~\cite[p.583]{Silberschatz} leveraging the MatchPy pattern matching library~\cite{krebber2018}, which takes into account the associativity of operators.}
\nic{Since, as mentioned in Section~\ref{sec:tr_ops}, we may instantiate transformers in many ways, for instance using different weighting models or other static properties, compiled pipelines can implement multiple transformers and/or operators, and can optimise their runtime execution depending on the IR software platform, as we illustrate in the following examples.}

\paragraph{Dynamic pruning optimisations.} \nic{By use of the rank cutoff operator (\%), typically we are applying a rank cutoff to a list of retrieved results generated by \retrieve{} transformer as a separate operation:}
\begin{equation}\begin{aligned} \label{eqn:topkexample}
    \textsf{top10} = \retrieve(index, \textsf{"BM25"}) \textsf{ \% } 10.
\end{aligned}\end{equation}

\nic{However, retrieval systems based on dynamic pruning techniques – such as MaxScore, WAND or BlockMaxWAND – can process queries and retrieve results faster when the number of documents to retrieve are reduced~\cite{INR-057}, as the higher scoring threshold means that more documents can be pruned, i.e., skipped over during retrieval. Hence the previous two-steps transformer can be automatically compiled as follows:}
\begin{equation*}\begin{aligned}
    \textsf{top10} = \textsf{Anserini}\retrieve(index, \textsf{"BM25"}, 10),
\end{aligned}\end{equation*}
\nic{where \textsf{AnseriniRetrieve} is an Anserini-specific implementation of the \retrieve{} transformer, which uses Lucene's BlockMaxWAND-based search backend~\cite{DBLP:conf/ecir/GrandMFL20}.}

\paragraph{Learning to rank optimisations.} \nic{To compute additional query-dependent features when re-ranking, the inverted index posting lists must be scanned until the requested docids are \crc{identified. This represents} a large computational overhead even if skipping is used~\cite{INR-057}.} \craig{Indeed, consider the following retrieval pipeline, which computes two additional query dependent features (query likelihood and TF.IDF)}:

\begin{equation}\begin{aligned} \label{eqn:features}
    &\textsf{first\_pass} = \retrieve(index, \textsf{"BM25"})\\
    &\textsf{tfidf} =  \retrieve(index, \textsf{"TFIDF"})\\
    &\textsf{ql} =  \retrieve(index, \textsf{"QL"})\\
    &\textsf{pipeline} = \textsf{first\_pass \textsf{>}\textsf{>} (tfidf \texttt{**} ql)}
\end{aligned}\end{equation}
\craig{As written, both additional query dependent retrieval features would result in additional access to the inverted index posting lists for each query.}

\nic{Instead, there have been two search architectures for computing additional query dependent features proposed in the literature: (1) the {\em doc vectors} approach~\cite{DBLP:journals/ir/AsadiL13}, where the direct index (which records the terms occurring in each document) is used for computing additional features, and (2) the {\em fat postings} approach~\cite{macdonald2013tois}, where the postings for the query terms of the documents that enter the final retrieved set are cached in main memory, allowing the fast computation of additional query-dependent features for the documents in the retrieved set.}
\nic{In both cases, when executing feature retrieval pipelines involving multiple query-dependent features, instead of computing the features one by one with multiple passes of the doc vectors/fat postings, the pipeline can be rewritten to a single Terrier \crc{retrieval} operation that extracts the fat postings first, \crc{then computes} all other features on the fat postings, without executing two expensive retrieval transformers:}
\begin{align*}
    & \textsf{pipeline} = \textsf{FeatureRetrieve}( index, \textsf{"BM25"}, [\textsf{"TFIDF"}, \textsf{"QL"}]).
\end{align*}
In the next section, we demonstrate the efficiency benefits of using an optimised pipeline upon retrieval operations involving two underlying IR platforms.

\pageenlarge{2}\section{Experiments}\label{sec:experiments}

The aim of the following experiments is to show that the optimisation of retrieval pipelines, as proposed in Section~\ref{sec:rewriting}, \craigcrc{results} in more efficient search executions. These experiments demonstrate that we can implement equivalent retrieval pipelines using multiple retrieval backends (Anserini and Terrier), which can be automatically optimised in different ways. In particular, we address the following research questions:
\begin{itemize}
    \item[RQ 1.] Does optimising the execution of rank cutoffs for dynamic pruning enhance the efficiency of an Anserini retrieval?
    \item[RQ 2.] Does optimising the execution of LTR to use the fat postings for computing multiple query dependent features enhance the efficiency of a Terrier retrieval?
\end{itemize}

\subsection{Experimental Setup}

We perform experiments on two standard TREC corpora, namely TREC Disks \crc{4\&5}, and ClueWeb09. We index each corpus using both Anserini and Terrier, while recording position information but otherwise using their default settings. This results in two indices for each corpus, containing 528,155 and 50,220,423 documents, respectively, for Disks 4\& 5, and ClueWeb09. For queries, we use corresponding TREC query sets: for \crc{Disks 4\&5}, 250 topics from the TREC Robust track '04, applying short (denoted T), medium (TD) and long (TDN) topic formulations; for ClueWeb09, we use 200 query-only topics from the TREC Web track 2009-2012.

Experiments are conducted on a Centos Linux 7.2 server with 96GB RAM and 12-core Intel E5-2609 CPUs. We use a single thread for all experiments, and report mean response time (MRT) in milliseconds. All experiments are conducted on our proposed framework \name{}.\footnote{The source code for our experiments \craigcrc{is at \url{https://github.com/} \url{cmacdonald/pyterrier_ictir2020}}.} \name is implemented in Python, and targets Anserini and Terrier backends, which are both written in Java -- indeed, we make use of the Pyjnius library\footnote{\url{https://github.com/kivy/pyjnius}} that permits easy interactions between Python and Java code.

\pageenlarge{2}\subsection{RQ1 Results}

In this experiment, we compare the response times of Terrier and Anserini in implementing the retrieval pipeline contained in Eq.~\eqref{eqn:topkexample}.
In particular, as Anserini uses Lucene's search engine backend based on BlockMaxWAND, the pipeline can be optimised. We further measure the response times of Terrier -- which does not deploy any dynamic pruning techniques -- for comparison. The top half of Table~\ref{tab:RQ12} provides the mean response times for the Terrier pipeline, and the unoptimised and optimised Anserini pipelines, on both the TREC Robust '04 and ClueWeb09 test collections.

From the results, it is apparent that for short title-only (T) queries on the Robust '04 index, Anserini benefits from use of the BlockMaxWAND dynamic pruning technique, although the benefits are less apparent for longer queries (TD and TDN), where Terrier outperforms Anserini. This is expected, as document-at-a-time dynamic pruning techniques such as BlockMaxWAND are known to be less efficient for longer queries~\cite{INR-057}.

Next, comparing Anserini Optimised and the Anserini Original retrieval pipelines, we see that by informing the Anserini backend of the number of documents required is 10 rather than 1000, mean response times are improved by up to 95\% on Robust '04 (short queries) and 63\% on ClueWeb09. Therefore, we conclude that, in answer to RQ1, applying the rank cutoff optimisation within the framework can \craigcrc{result} in marked efficiency \craigcrc{benefits for researchers}.

\begin{table}
    \caption{Mean response time (MRT, in milliseconds) of Terrier and Anserini for RQs 1 \& 2. MRTs are shown before (denoted Orig.) and after optimisation by rewriting (denoted Opt.). $\Delta$ is the \% improvement between original \& optimised.} 
\resizebox{85mm}{!}{
\centering
    \begin{tabular}{ccccccccc}
    \toprule
    Formulation & \multicolumn{2}{c}{Robust'04 T}  & \multicolumn{2}{c}{Robust'04 TD}  & \multicolumn{2}{c}{Robust'04 TDN} & \multicolumn{2}{c}{ClueWeb09} \\
     & MRT & $\Delta$& MRT & $\Delta$& MRT & $\Delta$ & MRT & $\Delta$\\
    \midrule
    \multicolumn{9}{c}{RQ1 - Rank Cutoff Optimisation} \\
    \midrule
        Terrier & 135.5 & - & 151.1 &- & 314.7 &- &1040.4 & - \\
        Anserini Orig. & 106.1 & - & 173.9 & - & 365.1 & - & 292.4 & - \\
        Anserini Opt. & {\bf 4.95} & -95\% & {\bf 24.65} & -85.8\% & {\bf 104.85} & -28.7\% & {\bf 107.4} & -63\% \\
\midrule
    \multicolumn{9}{c}{RQ2 - Learning to Rank Optimisation} \\
    \midrule
Anserini & 1336.1 & - & 1740.2 & - & 2625.4 & - & 3101.2 & -\\
        Terrier Orig. & 501.4 & - & 626.7 &  & 1032.9 & -& 2047.9 & -\\
        Terrier Opt. & {\bf 33.8} &  -93\% & {\bf 116.1} & -81\% & {\bf 350.8} & -66\% & {\bf 1255.8} & -39\% \\
    \bottomrule
    \end{tabular}   }
    \label{tab:RQ12} \vspace{-\baselineskip}
\end{table}


\subsection{RQ2 Results}
Next, we experiment to evaluate the efficiency of Anserini and Terrier in executing the retrieval pipeline contained in Eq.~\eqref{eqn:features}, i.e., retrieving a candidate set of documents for each query using BM25, before calculating additional two query dependent features, namely  TF.IDF and query likelihood with Dirichlet smoothing. Further, recall that such a retrieval pipeline can be optimised for Terrier, using the fat framework~\cite{macdonald2013tois} for calculating multiple query dependent features. The bottom half of Table~\ref{tab:RQ12} reports the resulting response times for the Anserini execution as well as the un-optimised and optimised Terrier executions.


\pageenlarge{2} From the table, we note that the Terrier implementation is faster at executing this complex retrieval pipeline than Anserini. This is expected -- at the time of writing, Anserini's reranking implementation uses one ``query'' to the underlying Lucene index for every document in the candidate set being re-scored. 
On the other hand, Terrier's original implementation is faster, but still requires one backend retrieval operation for each pipeline component. Finally, the optimised formulation only requires one backend fat \crc{retrieval} operation for each query. In doing so, this formulation makes use of the fat framework to allow the efficient calculations of multiple query dependent features in a single pass of the query term's posting lists. Finally, we note that the benefit of the fat framework decreases as the length of the queries increase (T $\rightarrow$ TD $\rightarrow$ TDN); this implies that the overheads of keeping the postings around for lots of query terms causes additional memory pressures. The alternative doc vectors approach~\cite{DBLP:journals/ir/AsadiL13} may be more efficient in such situations. Therefore, in answer to RQ2, we find that automatic optimisation of retrieval pipelines for LTR have the potential to markedly enhance the efficiency of such experiments.

Finally, it is worth emphasising that the point of this experiment is not to demonstrate a ``bake-off'', but instead to show that a single retrieval pipeline -- expressed in a conceptual manner -- can be executed on \name{} using multiple different retrieval backends. Moreover, that pipeline can be executed by those backends in different manners, with different efficiencies. The researcher need not be knowledgeable about the capabilities of those backends.


\section{Conclusions and Outlook}\label{sec:conclusions}

In this paper, we proposed a data model and framework for conducting IR experiments in a declarative manner. Our framework includes transformers representing standard retrieval operations, as well as operators for combining those transformers into retrieval pipelines. Further, we show how these pipelines can be automatically compiled and optimised, encoding knowledge of the capabilities of the underlying information retrieval system, to benefit the efficiency compared to semantically equivalent pipelines.

We believe that use of our framework can allow researchers to focus on creating transformers, for integrating their techniques with existing IR platforms -- such as Anserini or Terrier -- in end-to-end evaluation. The resulting code can be easily distributed as Jupyter notebooks, enhancing IR experiment reproducibility. In future, we believe that the proposed framework can be easily extended to support automatic parallelisation, by application of the pipeline using separate threads for different queries, as well as support for incremental querying, which would allow a neural re-ranker such as BERT to start training on some batches of training queries while the IR platform is still retrieving for further batches, rather than the current sequential executing of the retrieval pipeline.

\jeff{One thing that would make it stronger is making the use cases for when and how the formalisms gives us new / different capabilities. It feels like things are building to that, but it's not in the experiments, so it feels a bit missing perhaps.}

 \section*{Acknowledgements}
The authors would like to acknowledge Alex Tsolov who \crc{contributed to} the initial implementation of \name, along with other colleagues and contributors whose insights and contributions steered the \name development. \crc{Nicola Tonellotto was} partially supported by the Italian Ministry of Education and Research (MIUR) in the framework of the CrossLab project (Departments of Excellence).

\bibliography{bib}
\bibliographystyle{ACM-Reference-Format}
\end{document}